\documentstyle[twoside,fleqn,espcrc2]{article}
\newcommand{\be}{\begin{equation}}
\newcommand{\ee}{\end{equation}}
\newcommand{\AmS}{{\protect\the\textfont2
  A\kern-.1667em\lower.5ex\hbox{M}\kern-.125emS}}

\begin{document}
% declarations for front matter
\title{An improvement to the linear accept/reject algorithm}\thanks{
Presented by Lee Lin and Keh-Fei Liu at the conference}

\author{
L.~Lin\address{Department of Physics, National Chung Hsing University,
Taichung 40227, Taiwan, ROC},  K. F.~Liu
\address{Department of Physics and Astronomy,
University of Kentucky,
Lexington, KY 40506, USA} %\thanks{Speaker at the conference},
and
J.~Sloan$\mathrm{^b}$
}

%\begin{document}

\begin{abstract}
We study the improvement of Kennedy-Kuti's linear accept/reject algorithm
with different ordering criterion and modified Bhanot-Kennedy
estimator of $e^{\Delta H}$ to reduce probability-bound violation.
A new stochastic Monte Carlo algorithm to accommodate the
probability-bound violation is proposed.
\end{abstract}

\maketitle

\section{Introduction}                 \label{s1}

Sometime ago,  Kennedy and Kuti \cite{KK} proposed a Monte Carlo algorithm
which admits stochastically estimated transition probabilities as long as
they are unbiased. This opens up the door to solving problems where it is
feasible to estimate the transition probabilities but intractable
or impractical to calculate them exactly.

 The acceptance probability (denoted as $P_a$ from now on)
 in Kennedy-Kuti's linear algorithm is
\begin{eqnarray}
 & & {\rm P}_a(U_1\rightarrow U_2) = \lambda^+ \, +\,\lambda^-\,
 {\hat{e^{\Delta H}}}\, \nonumber \\
 &  & {\rm if}\,\,  f(U_1) > f(U_2)\,\, ,
\end{eqnarray}
\begin{eqnarray}
 & &{\rm P}_a(U_1\rightarrow U_2) = \lambda^- \, +\,\lambda^+\,
 {\hat{e^{\Delta H}}}\, \nonumber \\
 &  & {\rm if}\,\,  f(U_1)\le f(U_2)\,\, ,
\end{eqnarray}
where $\lambda^{\pm}$ are tunable real parameters ranging from 0 to 1,
${\hat r}$ denotes an unbiased
estimate of $r$ and $\Delta H = H(U_1)-H(U_2)$.
 $U_1$ denotes the old
configuration and $U_2$ the new (or proposed) configuration.
$f(U)$ is some observable of the gauge configuration U adopted for ordering
between $U_1$ and $U_2$.
Detailed balance can be proven to be satisfied.

But there is a major problem with the linear
algorithm. The value of
$P_a$ can violate the probability bound $[0,1]$ since it is estimated
stochastically. Once
 the probability bound restriction is not respected, detailed balance
is destroyed and systematic bias will show up. Hopefully, if the
bound violation occurs very rarely (e.g. once every one million times),
then the systematic bias might be very small and the expectation values
of various quantities
can still be correct within statistical errors \cite{KK}.

Within the framework of the
linear algorithm, there are at least three
ways to reduce the probability of bound violations:

(1) In general, the two tunable parameters $\lambda^\pm$ can be parameterized as
\be
\lambda^+ = 0.0\,\, ,\,\, \lambda^-={1\over 1+\epsilon}
\ee
where $0\le\epsilon<\infty$. So $ e^{\hat{\Delta H}} $
is allowed to fluctuate between
0 and $1+\epsilon$. If we increase $\epsilon$, then the allowed range
of $ e^{\hat{\Delta H}} $ will be larger and the probability 
of bound violations can be
reduced (although the intrinsic acceptance rate will be reduced
simultaneously).

(2) One can choose a better
 ordering criterion to reduce the
bound violation. One can imagine that when the ordering criterion is
uncorrelated with $\Delta H$,
then the upper bound will be violated quite frequently (about $50\%$ of the
time for $\lambda^- = 1$).  Thus, we believe the
 ``ideal" ordering criterion is  $\Delta H$ itself.
However, we cannot calculate $\Delta H$ exactly (otherwise
we would not need the stochastic estimator); the best we can do is to
estimate $\Delta H$ stochastically (without bias). We will obtain a
number (denoted as $x_0$) which can be made reasonably
close to the true value
of $\Delta H$, and use it as the ordering criterion. This should
greatly reduce the probability of upper-bound violations.

(3) Most importantly, one can reduce the variance of the estimated
acceptance probability either by brute force or by constructing
a better (unbiased) stochastic estimator. This will help  reduce
the probability of lower-bound violations
and further reduce upper-bound violations as well.

Although one can improve the performance of the linear algorithm with
these techniques, there are
still problems inherent to the algorithm which are impossible to
eradicate.
First of all, if we assume that the estimator of the acceptance probability
has a Gaussian distribution (which should be a reasonable assumption),
then no matter how hard we work, we will never
completely exclude the bound violations since the two long tails of
the Gaussian distribution always exist.
Secondly, the linear algorithm  with a stochastic estimator
is a volume-squared algorithm. This can be easily
seen by the following consideration. The variance of the estimated
acceptance probability is roughly proportional to $\delta^2/N$
where $\delta^2$ is the intrinsic variance and $N$ is the number of hits.
(This proportionality relation is true no matter what unbiased estimator
we choose.) The intrinsic variance $\delta^2$ is proportional to the
lattice size. Therefore, if we want to keep the bound violations under very
tight control, we need to work harder and harder (i.e. $N$ should become
larger and larger) as the lattice size grows. So $N$ should grow as the first
power of the volume $V$, and the cost will be proportional to $V^2$
since the cost of the stochastic estimator is usually proportional to V.
In real simulations on lattice QCD with dynamical fermions,
to lessen the probability bound violation in this way could be very costly.

In order to  completely
remove any systematic bias coming from bound violations,
and to reduce the cost of simulation
on larger lattices, we need to go beyond the linear
algorithm and come up with some new method. In the next
section, we propose
a new algorithm which will achieve these goals. We will see that
the new algorithm
eliminates the upper bound violation and absorbs the negative sign of the
lower bound violation into the observables
by introducing some auxiliary fields and
going back to the Metropolis accept/reject.

\section{The New Algorithm}

The first step toward the new algorithm is to transform the ``noise''
(coming from the stochastic estimator) into ``auxiliary fields", just
as pseudofermion fields are introduced as 
auxiliary fields. Therefore, the configuration
space is enlarged, and the updating and the accept/reject
will be carried out in this enlarged space. To be more precise, the partition
function $Z$ is
\begin{eqnarray}
Z &=&  \int [DU]\, e^{-S_G(U)}\, e^{-S_F(U)} \nonumber \\
  &=&  \int [DU][Dp]\, e^{-H(U,p)}\, e^{-S_F(U)} \nonumber \\
  &=& \int [DU][Dp][D\xi]P_\xi(\xi)\, e^{-H(U,p)}\, f(U,\xi),
\end{eqnarray}
where $U$ represents the original field (not necessarily the link variable),
$p$'s are the conjugate momenta introduced in the Hybrid Monte Carlo (HMC)
algorithm \cite{HBMC}
in case the molecular dynamics is used for updating. $H(U,p) = \frac{p^2}{2}
+ S_G(U)$ is the hamiltonian and $\xi$'s 
 are the auxiliary fields ``representing"
the stochastic fields. $P_\xi$ is the
probability distribution for $\xi$'s.
The function $f(U,\xi)$ is any function which satisfies
%Together with $f(U,\xi)$, they define the function $e^{-S_F(U)}$,
\be \label{fU}
e^{-S_F(U)}\,\, =\,\,\int [D\xi]\,P_\xi(\xi)\,
\, f(U,\xi),
\ee
\i.e. $f(U,\xi)$ is an unbiased estimate of $e^{-S_F(U)}$ for noise $\xi$.

This completes the
first step toward the new algorithm. The space is indeed enlarged.
Originally, we had a configuration space of the $U$'s. HMC enlarges the
space to $(U,p)$. Now it is further enlarged to
$(U,p,\xi)$ with the inclusion of the stochastic field $\xi$. 
This first step actually sets up the
platform for performing updates in the enlarged space.
So from now on, when we specify a state, we are specifying a state in this
enlarged space.

One should also notice that everything is general so far. We have not written
down a specific functional form for $f(U,\xi)$ yet,
nor have we decided what $P_\xi$ should be.
 So this means the
new algorithm can be applied to a general class of problems. It is not
restricted to a specific type of problem. The auxiliary fields can have
arbitrary distributions (e.g. Gaussian, $Z_2$ etc.)

The second step of this new algorithm is to observe that
\be
f(U,\xi)\,\, = \,\,{\rm sign}(f)\,|f(U,\xi)|\,\, .
\ee
Therefore ${\rm sign}(f)$, which means the sign of the function $f$,
is a state function. We then can do the following:
\begin{eqnarray}  \label{A}
\langle A \rangle &=& \int[DU][Dp][D\xi]\,P_\xi(\xi) \nonumber \\
 & & A(U)\,{\rm sign}(f)\,e^{-H(U,p)}\,|f(U,\xi)|,
\end{eqnarray}
where $A$ is the quantity we want to measure.
We now reinterpret $|f|$, which is positive definite, as part of the 
probability distribution. So we are measuring the quantity $A(U){\rm sign}(f)$
under the 
probability distribution $e^{-H(U,p)}\,|f(U,\eta,\rho)|$ and the 
partition function $Z = \langle {\rm sign}(f)\rangle$.
This in principle takes care of the lower probability-bound violations. 
(Actually they are is replaced with a potential sign problem.)
Notice that this reinterpretation is possible because the sign of
$f(U,\eta,\rho)$ is a state function. In the linear accept/reject,
one calculates the transition probability directly (and stochastically).
There, the transition probability is not a state function. When we
obtain a minus sign there,
it can not be swept into the observable as in Eq. (\ref{A}).

The third part
of the new algorithm is to go back to Metropolis to do the accept/reject.
There are two accept/reject steps. The first one is to propose
updating of $U$ and $p$ in the usual way, e.g. molecular dynamics evolution 
while keeping the
stochastic field $\xi$ fixed. The acceptance probability $P_a$ is
\begin{eqnarray}
 & &P_a(U_1,p_1,\xi \rightarrow U_2,p_2,\xi)\,\,= \nonumber \\
 & & {\rm min}\Bigl(1,{e^{-H(U_2,p_2)}\,|f(U_2,\xi)|\over
                 e^{-H(U_1,p_1)}\,|f(U_1,\xi)|}\Bigr)\,\,  .
\end{eqnarray}
The second accept/reject step refreshes the stochastic field $\xi$ according 
to the probability distribution $P_{\xi}(\xi)$ while keeping $U$ and $p$ fixed.
The acceptance probability in then
\begin{eqnarray}
 & & P_a(U,p,\xi_1\rightarrow U,p,\xi_2)\,\,= \nonumber \\
 &  & {\rm min}\Bigl(1,{|f(U,\xi_2)|\over
                  |f(U,\xi_1)|}\Bigr)\,\,  .
\end{eqnarray}
Obviously, there is no probability-bound violation in either of these Metropolis
accept/reject steps.

Detailed balance can be proven to be satisfied.
It is obvious that all the techniques for reducing the
variance of the estimator developed before can be applied here.

%Notice that the above $P_a$ is chosen for that case where $\eta$'s and
%$\rho$'s are kept fixed during the evolution (i.e. the process of
%proposing the new configuration). If $\eta$'s and $rho$'s are also updated,
%then we only have to slightly change $P_a$ to preserve detailed balance.
%then we should have
%$$
%P_a(U_1,p_1,\eta_1,\rho_1\rightarrow U_2,p_2,\eta_2,\rho_2)\,\,=
%$$
%$$
%{\rm min}\Bigl(1,{P_\eta(\eta_2)\,P_\rho(\rho_2)\,
%                  e^{-H_{pG}(U_2,p_2)}\,|f(U_2,\eta_2,\rho_2)|\over
%                  P_\eta(\eta_1)\,P_\rho(\rho_1)\,
%                 e^{-H_{pG}(U_1,p_1)}\,|f(U_1,\eta_1,\rho_1)|}\Bigr)\,\,  .
%$$

We have tested the new algorithm on a very simple 5-state model.
(This is the same model used in \cite{KK} for demonstration.)
We calculated the average energy with the linear algorithm and the new
algorithm. Some data are presented in table 1. (The exact value is
$0.180086$.)

\begin{table}[tb]
\caption{  \label{t1} Data of average energy
obtained by using linear and new algorithms. $\epsilon = 0$ in this case.}
\begin{center}
\begin{tabular}
{|c|r@{.}l|r@{.}l|}
\hline
{$\delta^2$}
&\multicolumn{2}{c|}{Linear} & \multicolumn{2}{c|}{New}
\\
\hline
1.0    &  0&18009(14) & 0&17994(14) \\
\hline
3.0    &  0&18127(14) & 0&17999(14) \\
\hline
6.0    &  0&18216(14) & 0&18014(14) \\
\hline
10.0   &  0&18257(14) & 0&18005(14) \\
\hline
20.0   &  0&18342(14) & 0&17995(14) \\
\hline
\end{tabular}
\end{center}
\end{table}
We see that as the intrinsic variance of the estimator grows, the linear
algorithm introduces a systematic error (mostly due to upper-bound
violations). The new algorithm, however, still gives the correct value within
errors.

To apply this new algorithm to the dynamical fermion problem, we
note that the fermion determinant can be calculated stochastically as
a random walk process~\cite{BK}
$$
e^{Tr\ln M} = 1 + Tr\ln M (1 + \frac{Tr \ln M}{2} (1 + \frac{Tr \ln M}{3}
(...,
$$
which can in turn be written in the following integral
\begin{eqnarray}
& & e^{Tr\ln M}=\int \prod_{i =1}^{\infty} {\cal D}\eta_i \,\delta(|\eta_i| -1)
     \int_{0}^1 \prod_{n =1}^{\infty} {\cal D} \rho_n \nonumber \\
& &  [1 + \eta_1^{\dagger} \ln M \eta_1
  (1 + \theta(\rho_2 - 1/2) \eta_2^{\dagger} \ln M \eta_2  \nonumber \\
& &  (1 + \theta(\rho_3 - 2/3) \eta_3^{\dagger} \ln M \eta_3 (......]  \nonumber
\end{eqnarray}
This sequence terminates stochastically in finite time and only the
seeds from the pseudo-random number generator need to be stored in practice.
Comparing this to Eq. (\ref{fU}), the function $f(U,\eta,\rho)$ (the $\xi$ in
Eq. (\ref{fU}) is represented by 
two stochastic fields $\eta$ and $\rho$ here) can be defined for the
fermion determinant. One can then use the efficient Pad\'{e}-Z$_2$
algorithm~\cite{TDLY} to calculate $f(U,\eta,\rho)$. 

While we are in the process of applying this algorithm to QCD, we have
considered how efficiently the HMC with pseudofermion is as far as updating
$Tr\ln M$ is concerned. To this end, we have run the HMC with
pseudofermions on an $8^3 \times 18$ lattice at $\beta = 5.6$ and
$\kappa = 0.154$. 40 steps are taken with step size of 0.025 in 170
molecular dynamics trajectories. We calculated the correlation between
the change of $Tr \ln M$ and the change of the pseudofermion action
$S_F$ along a trajectory. The latter is designed to emulate the former. We 
find the normalized correlation is
\begin{equation}
\langle \Delta Tr\ln M \,\,\Delta S_F\rangle = %\color{red}
{0.053 \pm 0.133},
\end{equation}
which is quite small. This raises a question as to how efficient
the HMC with pseudofermions might be in updating the fermion determinant.
Further study of the autocorrelation of $Tr \ln M $ is needed to
answer this question.

\section{Summary and Discussion}                              \label{s5}

In summary, the new algorithm solves the problem of probability-bound 
violations which has been troubling the linear accept/reject algorithm.
The upper-bound violation problem is solved by going back to the Metropolis
accept/reject (in the enlarged space). The lower-bound violation problem is
solved by grouping the sign with the observable.
With the probability bound violation problem solved, we do not
have to have an extremely small variance in the stochastic estimation.
So, in principle, the volume-squared
behaviour of the algorithm is less sever and hopefully can be put under
control.
%%\vspace{3pt}
%%%%%%%%%%%%%%%%%%%%%%%%%%%%%%%%%%%%%%%%%%%%%%%%%%%%%%%%%%%%%%%%%%%%%%%

\section{Acknowledgement}

The authors would like to thank A. Kennedy for stimulating discussions and
for helping clarify the Metropolis steps. 
This work is partially support by the U.S. DOE grant DE-FG05-84ER40154.
Lee Lin wishes to thank NSC of the Republic of China
for support under the contract number 36083F.

\end{document}